\documentclass[journal=jacsat,manuscript=article]{achemso}

\usepackage[version=3]{mhchem} % Formula subscripts using \ce{}

\author{Mathieu Massicotte}
\email{mathieu.massicotte@mail.mcgill.ca}
\author{Victor Yu}
\author{Eric Whiteway}
\author{Dan Vatnik}
\author{Michael Hilke}
\affiliation[McGill]
{Department of Physics, McGill University, Montr\'eal, Qu\'ebec, H3A 2T8, Canada}

\title[An \textsf{achemso} demo]
  {Quantum Hall Effect in Fractal Graphene: growth and properties of graphlocons}


\begin{document}
%%%%%%%%%%%%%%%%%%%%%%%%%%%%%%%%%%%%%%%%%%%%%%%%%%%%%%%%%%%%%%%%%%%%%
%% The manuscript does not need to include \maketitle, which is
%% executed automatically.  The document should begin with an
%% abstract, if appropriate.  If one is given and should not be, the
%% contents will be gobbled.
%%%%%%%%%%%%%%%%%%%%%%%%%%%%%%%%%%%%%%%%%%%%%%%%%%%%%%%%%%%%%%%%%%%%%
\begin{abstract}
Highly dendritic graphene crystals up to 0.25 mm in diameter are synthesized by low pressure chemical vapor deposition inside a copper enclosure. With their sixfold symmetry and fractal-like shape, the crystals resemble snowflakes. The evolution of the dendritic growth features is investigated for different growth conditions and surface diffusion is found to be the growth-limiting step responsible for the formation of dendrites. The electronic properties of the dendritic crystals are examined down to sub-Kelvin temperatures, showing a mobility of up to 6300 cm$^2$V$^{-1}$s$^{-1}$ and quantum Hall oscillations are observed above 4T. These results demonstrate the high quality of the transport properties despite their rough dendritic edges.
\end{abstract}


The advent of large-scale graphene grown by chemical vapor deposition (CVD) on transition metals opens a viable and promising route towards the commercialization of graphene-based electronics.\cite{Li2009e,Bae2010, Novoselov2012} The growth of graphene on copper has attracted considerable interest due to the simplicity, scalability, affordability, and homogeneity of the synthesized film. While this method solves the obvious problem of small-scale production associated with exfoliated graphene, it often results in a graphene film with lower electronic performance.\cite{Cao2010,Li2009e} Significant efforts have been made to reduce the extrinsic performance-limiting factors such as chemical impurities and structural damages.\cite{Annealing2011,Liang2011,Pirkle2011,Chan2012} More recently, several studies have focused on improving the intrinsic electrical properties of CVD graphene. In particular, theoretical and experimental works have identified grain boundaries as one of the main sources of disorder in CVD graphene films.\cite{Yazyev2010,Huang2011,Yu2011,Graphene2011a}

Two approaches have been considered to overcome this problem. One consists in improving the electronic transport through the grain boundaries by engineering the growth conditions.\cite{Tsen2012} The second strategy aims at decreasing the number of nucleation sites and increasing domain size in order to reduce the impact of grain boundaries on the electrical properties of the film. Following the pioneering work by Li et al. \cite{Li2011a}, several CVD processes have been proposed\cite{Fan2011a,Zhang2012a,Wang2012} to grow large crystals with lateral lengths up to 2.3 mm.\cite{Graphene2012} These crystals display various morphologies: hexagons, flowers, squares and dendritic hexagons. Here, we report the growth of large, highly dendritic graphene crystals which we dubbed \textit{graphlocons} due to their resemblance to snowflakes. Monolayer graphlocons, up to 250 $\mu$m in lateral size with very few defects were grown. We compared their growth shape evolution with other large island growths in order to confirm their unique morphology and propose a mechanism for the formation of dendrites. Field-effect transistors (FETs) were fabricated on SiO$_2$/Si based on graphlocons and field-effect mobilities up to 6300 cm$^2$/V$^{-1}$s$^{-1}$ were measured at 4 K. These devices also displayed well-developed quantum Hall effect (QHE) features despite their dendritic edges.

\section{Results and discussion}

\begin{figure}
 \centering
\includegraphics[width=\columnwidth]{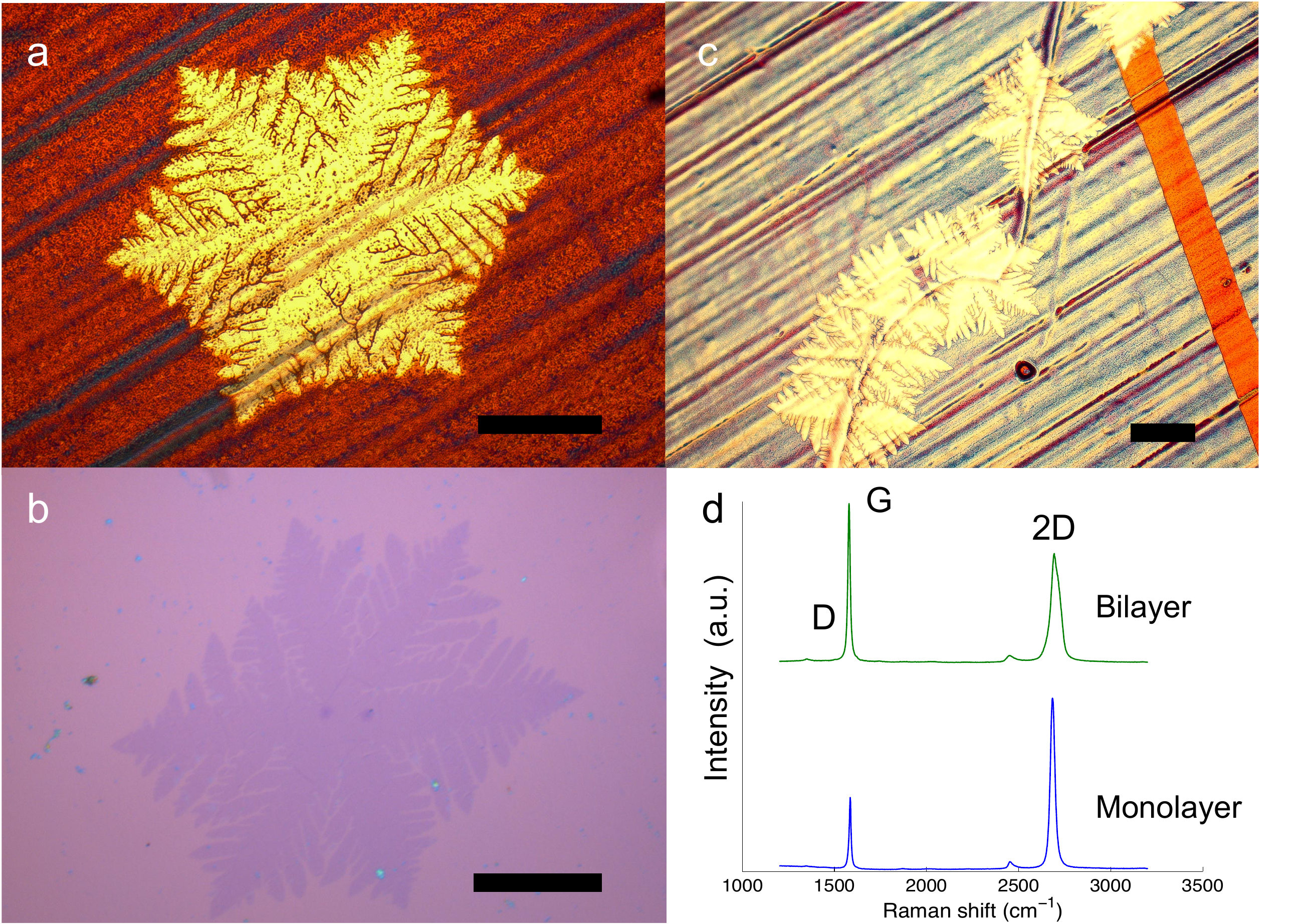}
  \caption{(a,b) Optical micrograph of as-grown graphlocons inside a Cu enclosure. The dark colored region corresponds to oxidized Cu.(c) Optical micrograph of a graphlocon transferred to a SiO$_2$/Si substrate and (d) Raman spectra taken on a branch (monolayer) and center (bilayer) of a graphlocon. The scale bars on (a,b,c) are 50 $\mu m$.}
  \label{fgr:synthesis}
\end{figure}

Graphlocons were synthesized inside a copper-foil enclosure by employing a technique similar to the one reported by Li et al.\cite{Li2011a}, but using a vertical quartz tube and higher gas pressure. The Cu enclosure was first annealed at 1025$^{\circ}$C for 30 min in 150 mTorr of H$_2$ flowing at 3 sccm.  The growth was performed for 30 mins at 1025$^{\circ}$C at a pressure of 1500 mTorr, using a 0.5 sccm CH$_4$ flow and a 3 sccm H$_2$  flow. Once the growth is completed, graphlocons are visible optically by heating up the copper foil in air on a hot plate for about 2 min at 200$^{\circ}$C. This simple procedure results in the oxidation of the copper regions which are not covered with graphene, creating a high contrast with those protected by graphene.\cite{Chen2011}  Using an optical microscope, most graphlocons appear as bright six-fold snowflakes over the colored polycrystalline copper substrate (\ref{fgr:synthesis}a). The diameter of these domains varies between a few microns and 250 $\mu$m.  Four-fold islands were also found in specific regions, indicating that the growth might be affected by the crystal orientation or morphology of the underlying copper substrate.\cite{Wu2011b,Wofford2010} Moreover, we observed a much higher domain density in regions of rough copper surface. \ref{fgr:synthesis}b, which shows the growth of graphlocons along a pre-existing scratch, demonstrates clearly that the copper morphology has an effect on the nucleation behavior.\cite{Han2011}

The seeding role of copper impurities was also indicated by the presence of bilayer (and few-layer) graphene at the center of some domains. These bilayer structures are easily observable once the domains are transferred onto a SiO$_2$/Si substrate. As \ref{fgr:synthesis}c indicates, more than one of those terraced structures could sometimes be seen in the central region, suggesting that graphlocons can be composed of more than one crystal. To confirm the presence of multilayer graphene, Raman spectra were taken on monolayer and multilayer regions of a graphlocon. The bottom spectrum of \ref{fgr:synthesis}d was measured with the laser aiming in the middle of one of the branches. It corresponds to a graphene monolayer, with a 2D to G peak intensity ratio of I$_{2D}$/I$_G\approx2$ and a 2D-peak FWHM of 35 cm$^{-1}$. The top spectrum  was obtained by directing the laser on a darker central region. It yields a much smaller I$_{2D}$/I$_G$ ratio ($\sim$0.7) and a 2D-peak twice as broad (FWHM $=63$ cm$^{-1}$), which indicates the presence of a bilayer/multilayer in the center of the graphlocon.\cite{Ferrari2006} We also notice that in both spectra the defect-induced D-peak is very weak, indicating the high quality of the graphlocon.

%\subsection{Growth Mechanism}

\begin{figure}
 \centering
\includegraphics[width=\columnwidth]{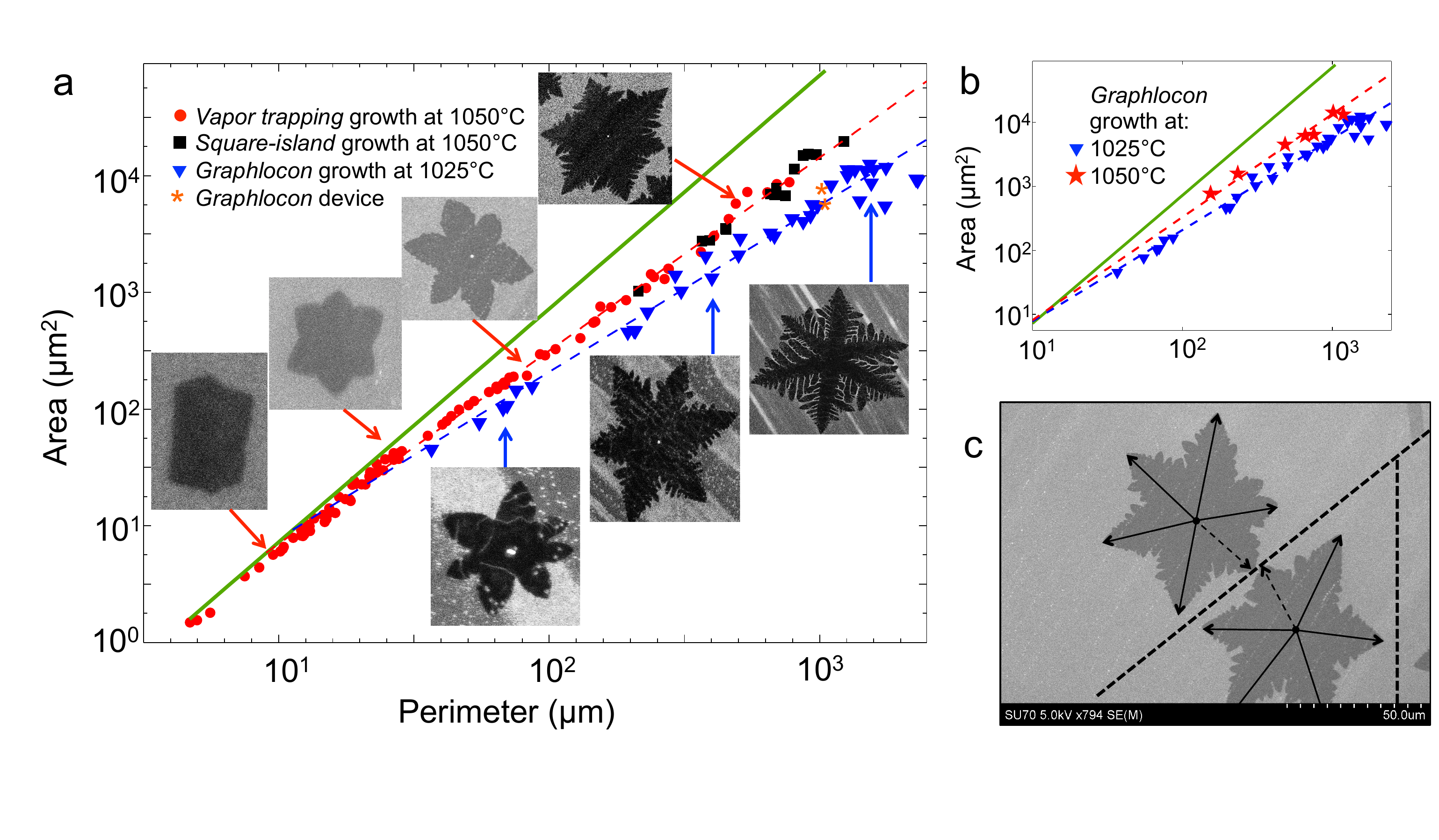}
  \caption{Shape evolution of graphlocons. (a,b) Plots of log (island area) as a function log (island perimeter) for various growth conditions and techniques. Each data point represents an individual island and SEM images are shown in (a) for some of them. Islands represented by red dots and black squares were grown using conditions similar to Ref.~\citenum{Zhang2012a} and Ref.~\citenum{Wang2012}, respectively. The dotted lines are linear fits of data points with perimeter > 10 $\mu m$ and the solid line corresponds to the behavior expected for perfect hexagonal (i.e. non-dendritic) islands. c) SEM image showing the effect of competitive capture on the graphlocon shape. The dotted line illustrate the capture zones and the arrows highlight the anisotropy of the growth.}
 \label{fgr:growthmech}
\end{figure}

To highlight and quantify the distinct morphology of the graphlocons, we compared them to graphene islands resulting from other CVD methods employed for growing large crystals. Flower-shaped crystals were obtained using the vapor trapping method described by Zhang et al.\cite{Zhang2012a} and square-shaped islands were grown using conditions similar to those reported by Wang et al.\cite{Wang2012}  (see Supporting Information for the details of the growth). \ref{fgr:growthmech}a shows a log-log plot of the area  (A) of individual islands versus their perimeter (P), as measured from SEM images. Data from all growth methods are included and the dashed lines associated with each type of growth are linear fits. The solid line corresponds to the relationship between perimeter and area for perfect hexagons. For all growth methods, islands grow with a scaling exponent $\alpha$ < 2 ($A\sim P^\alpha$), as expected for branched or fractal growth like diffusion-limited aggregates.\cite{Scott2006,Fairbanks2011} The fact that $\alpha$ is lower for graphlocons ($\alpha$ = 1.43) than for other types of growths ($\alpha$ = 1.66) clearly demonstrate their higher degree of ramification.

All growths presented in \ref{fgr:growthmech}a show a similar island shape evolution, which suggests that a single growth mechanism could be at work. A possible explanation was first proposed by Nie et al.\cite{Nie2011} who argued that graphene on Cu(111) is surface diffusion limited. In this growth regime, the shape of the island stems from two competitive processes: (1) carbon atoms or aggregates attach to the island boundary at a rate $k = 1/\Delta t$, where $\Delta t$ is the time between two atom impingements. (2) Atoms diffuse or detach/reattach along the island edge in order to preserve the thermodynamic shape of the graphene island. Assuming a random diffusion process, the time $t_d$ needed for an atom to diffuse along a boundary of size $L$ is $t_d\propto L^2/D$, where $D$ is a diffusion coefficient. When $\Delta t > t_d$, atoms have enough time to diffuse before a new impingement occurs and the equilibrium shape dominates. This shape, which minimizes the edge free energy, can be found by the Wulff construction\cite{Stability2012, Gao2012} and corresponds to a compact hexagon with zigzag edges, as reported by several experiments.\cite{Luo2011a,Yu2011,Geng2012} As the island grows, the diffusion time increases and faster growing orientations start ``growing out'' when $t_d > \Delta t$. % (and the crystal grows asymptotically toward its kinetic Wulff shape \cite{Sekerka2005}, which is proportional to the edge free energy \cite{Artyukhov2012}).%
For graphene on Cu(111), the growth rate has a six-fold, ``flower-like'' symmetry, with slow and fast growing orientation corresponding to zigzag and armchair edges, respectively.\cite{ Artyukhov2012,Nie2011,Stability2012} In this growth regime, dendrites arise from Mullins-Sekerka\cite{Mullins1963} type shape instabilities and grow along faster growth orientations.

A transition from compact to ramified morphologies can be seen in \ref{fgr:growthmech}a for graphene grown with the vapor trapping method. It corresponds to the point where the curve deviates from the hexagonal geometry ($\alpha$ = 2) and starts following a dendritic growth ($\alpha$ < 2). This transition occurs when $t_d = \Delta t$,  which defines the correlation length $L_c$ of the dendrites. A transition was not observed for the other types of growth due to the limited range of island sizes we synthesized. Interestingly, the extrapolated transition point of the graphlocon growth curve coincides with the vapor trapping one. Past this transition point, six-fold branches form and dendritic arms progressively grow on their sides. For large graphlocons, the development of secondary dendrites can be observed on primary dendrites, thus illustrating the self-similar nature of this growth. We note that dendrites grow preferentially with a 60$^{\circ}$ angle with respect to their parent dendrite (or branch), consistently with a six-fold growth symmetry. %New part. %
According to our interpretation of the growth, we should also expect a change in the shape evolution with the growth temperature since it affects the surface diffusion of carbon species ($D$ can be described by an Arrhenius equation \cite{Roder1995}). As \ref{fgr:growthmech}b indicates, growing graphlocons in the same conditions but at higher temperature changes the value of $\alpha$ from 1.43 (T = 1025$^{\circ}$C) to 1.61 (T = 1050$^{\circ}$C), which is consistent with an increase in $D$.  Additional growth experiments show that dendrites can be suppressed at higher growth temperature (see Supporting Information). Furthermore, we observe that the island shape is affected by the proximity of neighboring islands such that branches tend to grow longer toward regions of low island density (\ref{fgr:growthmech}c). This is a clear hallmark of competitive capture between islands sharing the same diffusion field.\cite{Nie2011,Scott2006} This competition for the capture of the same carbon species alters the capture zone of each island and results in an asymmetric growth rate. All aforementioned observations provide evidence that the growths investigated are surface diffusion limited.

%\subsection{Electronic properties}

\begin{figure}
 \centering
\includegraphics[width=12cm]{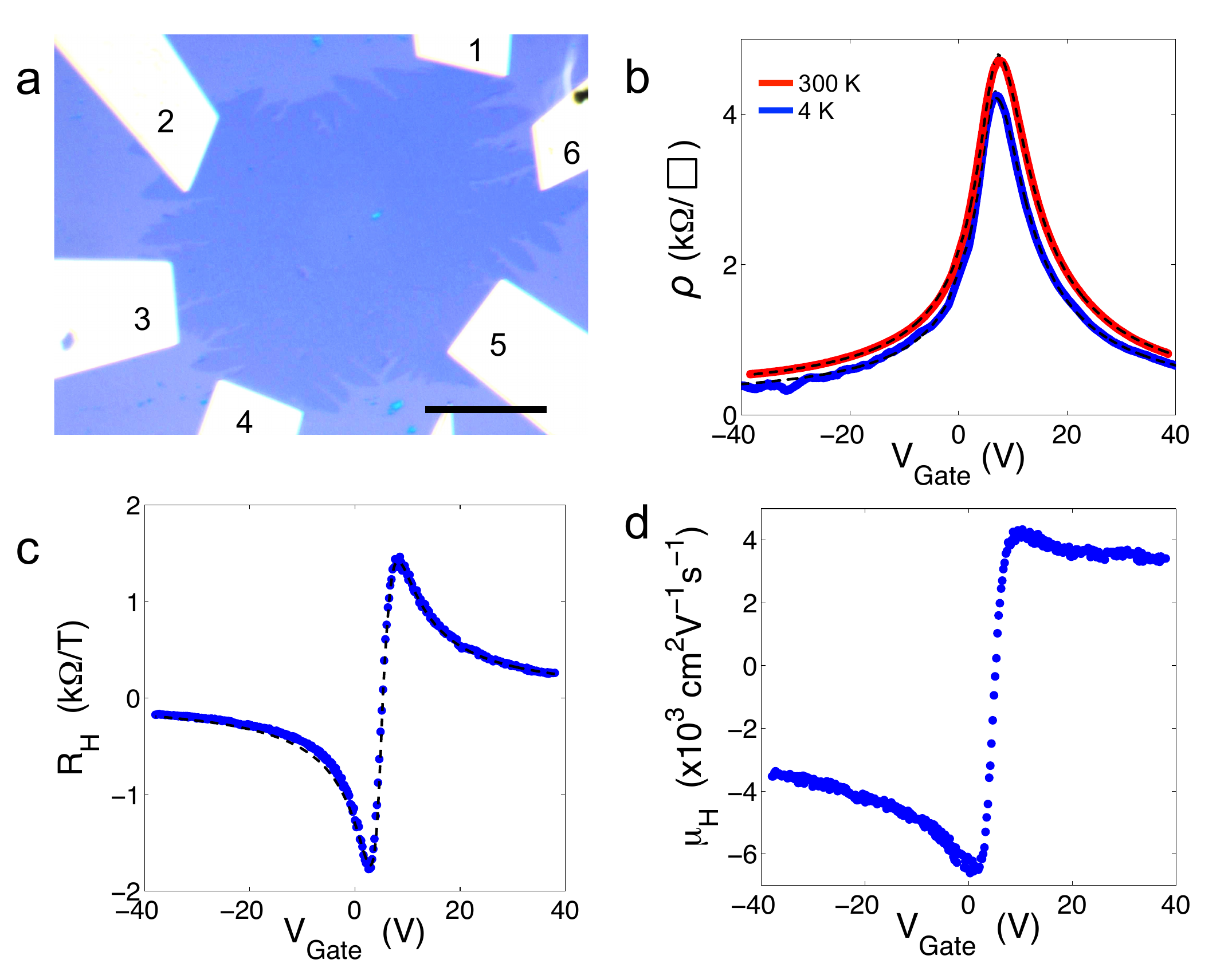}
  \caption{Electronic transport of a graphlocon FET on SiO$_2$/Si. (a) Optical micrograph of the device. The scale bar is 30 $\mu m$. (b) Sheet resistance as a function of gate voltage at 300 K (red) and 4 K (blue). (c) Hall resistance and (d) Hall mobility as a function gate voltage at 0.3K. The dotted lines are fits to the data using a diffusive transport model. }
  \label{fgr:electronic}
\end{figure}

To assess the effect of dendrites on the electronic transport properties of graphlocons, we transferred them onto a SiO$_2$ /Si substrate and electrically contacted their lobes to make a back-gated graphene FET.  Two such devices were cooled down to 300 mK in a pumped $^3$He refrigerator and magnetotransport measurements yielded similar results. Their morphological features are indicated in \ref{fgr:growthmech}a. In what follows we only present data for one of them, shown in \ref{fgr:electronic}a. The sheet resistance was obtained by the Van der Pauw (VdP) method \cite{VanDerPauw1958} using the leads 1, 2, 3 and 5.

\ref{fgr:electronic}b shows the change in sheet resistance ($\rho $) as a function of gate voltage ($V_{G}$) at 300 and 4 K without magnetic field. During cool down, the position of the charge neutrality point ($V_{Dirac}$) shifted slightly from 7 V to 5.2 V and the overall resistivity decreases by 10\%, indicating a metallic behavior. Both curves display an on/off ratio of $\sim$9 within the gate voltage range displayed in \ref{fgr:electronic}a. To fit these curves we used a diffusive transport model similar to the one proposed by Morozov et al. \cite{Morozov2008}. They showed that the inverse of the resistivity, after a contribution due to the short-range scattering $\rho_S$ is subtracted, depends linearly with gate voltage $(\rho-\rho_S)^{-1}\simeq \mu C_{ox}(V_G-V_D)+\sigma_{res}$, where $\mu$ is the field effect mobility, $C_{ox}$ the gate capacitance and $\sigma_{res}$ the extrapolated residual conductivity at the charge neutrality point. However, in order to account for the difference in mobilities for holes and electrons as well as the existence of a residual density $n_0$ at the charge neutrality point due to large scale inhomogeneities, we can write the total carrier density as $n+p=\sqrt{(n-p)^2+n_0^2}$ and $n-p=C_{ox}(V_G-V_{Dirac})/e$, where $n$ and $p$ are the densities of electrons and holes, respectively \cite{Dorgan2010}. This leads to
\begin{equation}
\rho=\frac{1}{e(\mu_nn+\mu_pp)}+ \rho_S,
\end{equation}

\noindent with $C_{ox}$ = 11.5 nF/cm. Using this equation, we extracted an electron mobility $\mu_n\simeq $4300 cm$^2$V$^{-1}$s$^{-1}$ and $\mu_p\simeq$6300 cm$^2$V$^{-1}$s$^{-1}$ for holes at 4K. The residual density was found to be $n_0\simeq3.9\times10^{11}$ cm$^{-2}$ and the short-range scattering resistivity $\rho_S\simeq $105 $\Omega$. These values compare well with those commonly measured in exfoliated graphene \cite{Dorgan2010}.  For a more thorough investigation of the carrier mobility, we also measured the Hall resistance R$_H$ at low B-field (\ref{fgr:electronic}c) and extracted the Hall mobility $\mu_H$ = $R_H/\rho$ (\ref{fgr:electronic}d). The gate voltage dependence of $R_H$ for two carriers is given by
\begin{equation}
R_H=\frac{(p-n)}{(n-p)^2+n_0^2}
\end{equation}
\noindent which agrees well with our measurements as shown in \ref{fgr:electronic}c. The Hall mobility was found to vary significantly as a function of $V_G$, especially in the hole doped regime. The highest values of $\mu_H$ for holes and electrons match those of the field effect mobility $\mu$ defined above.

\begin{figure}
 \centering
\includegraphics[width=\columnwidth]{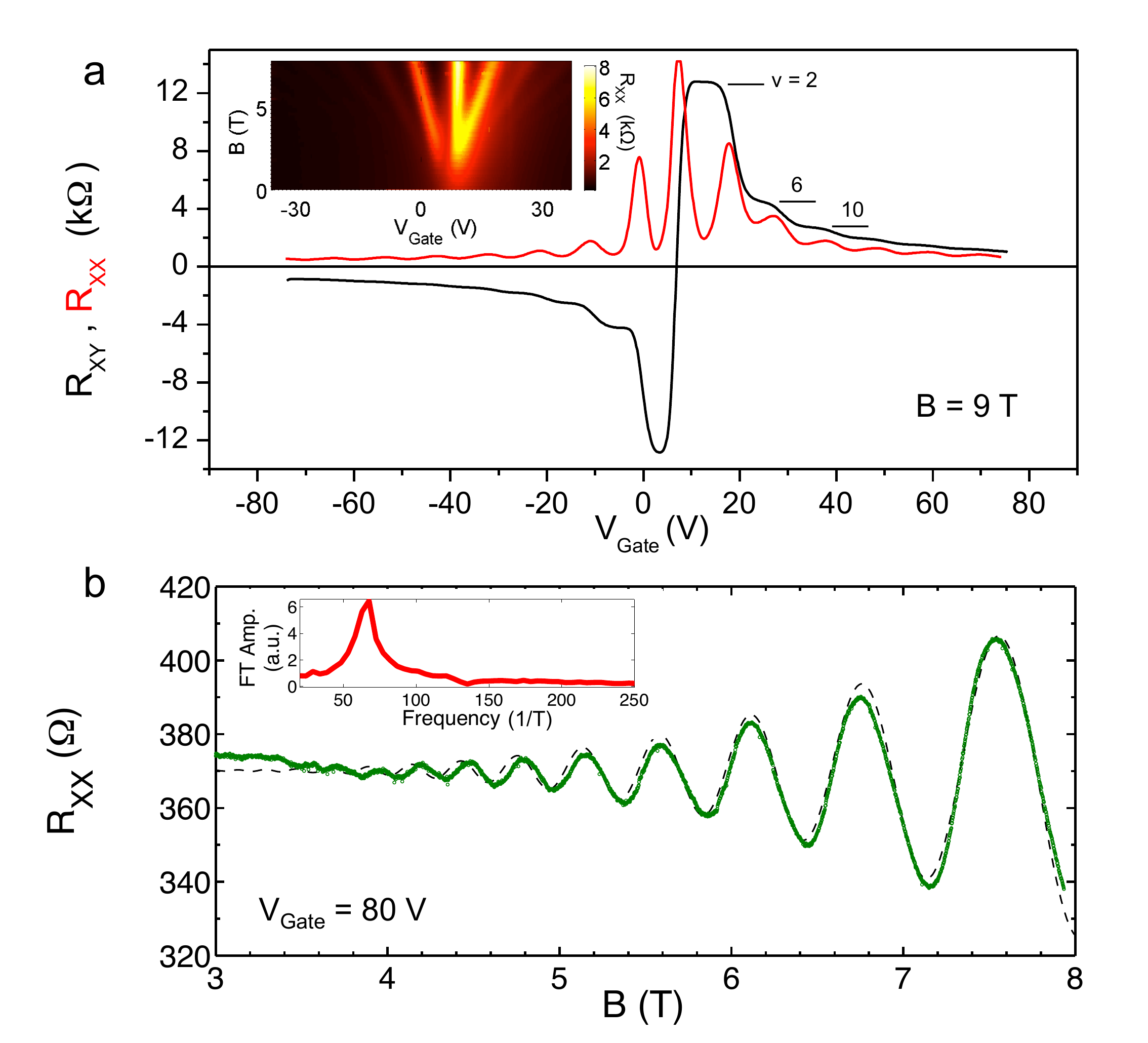}
  \caption{Magnetotransport of a graphlocon FET on SiO$_2$/Si. (a) Hall resistance $R_{xy}$ (black line) and longitudinal resistance $R_{xx}$ (red line) as a function of gate voltage $V_{Gate}$ at B = 9 T and T = 0.3 K. The quantum Hall plateaus at $\nu$ = 2, 6 and 10 are indicated. The inset shows the Laudau fan diagram of  $R_{xx}$  as a function of gate voltage and perpendicular magnetic field. (b) $R_{xx}$ as a function of B measured at $V_{Gate}$ = 80 V. The dotted line is a fit for the SdHO. The inset shows the Fourier transform of $R_{xx}$ vs $B^{-1}$ and the peak represents the charge carrier density up to a factor 4$e/h$.}
  \label{fgr:magneto}
\end{figure}

The homogeneity and quality of the graphlocon sample is also reflected by the magnetotranport measurements which display clear quantum Hall physics. \ref{fgr:magneto}a shows the longitudinal ($R_{xx}$) and Hall resistivity ($R_{xy}$) as a function of $V_{G}$ measured in a perpendicular magnetic field B = 9 T and T = 0.3 K in the sample shown in \ref{fgr:electronic}a. $R_{xx}$ and $R_{xy}$ were obtained by passing a small, low frequency current through contacts 3-5, and measuring the voltage between contact 1-2 and 2-4, respectively. $R_{xx}$ was multiplied by a geometrical factor of 4.5 as derived from the VdP method. The data shows clear quantum Hall features, with well-resolved Hall plateaus and deep minima of $R_{xx}$ at filling factors $\nu$ = $\pm$2, $\pm$6 and $\pm$10 as expected for monolayer graphene. In the inset of \ref{fgr:magneto}a, we show $R_{xx}$ as a function of gate voltage and magnetic field. The resulting Laudau fan diagram shows the emergence of the quantum Hall states. Quantized Landau levels appear as maxima lines coming out of B = 0 T and their linear dependence in B and $V_G$ agree with the behavior for monolayer graphene. \cite{Sarma2010}

Finally, we investigated the Shubnikov-de Hass oscillation (SdHO) displayed by $R_{xx}$ at high gate voltage ($V_G$ = 80 V) for B > 3.4 T, as shown in \ref{fgr:magneto}b. The figure's inset shows the Fourier transform of $R_{xx}$  as a function of B$^{-1}$ which displays a prominent peak at 62$\pm$5 T$^{-1}$. This value corresponds to the carrier density up to a factor 4$e/h$, yielding $n$ = (6.0$\pm$0.5)$\times10^{12}$ cm$^{-2}$ which is consistent with $V_G$-$V_{Dirac}$=79$\pm$6 V. Using the expression detailed by Babinski et al. \cite{Babin2000}, the SdHO allows us to extract a quantum mobility $\mu_Q$ of 1100 cm$^2$V$^{-1}$s$^{-1}$ with the fit shown in  \ref{fgr:magneto}b. The extracted quantum mobility is about five times smaller than the Hall and field effect mobilities and characterizes the effective broadening of the Landau levels due to disorder. Our observed magnetotransport features are comparable to those observed for typical exfoliated graphene samples\cite{Dorgan2010} and high-quality CVD-graphene samples \cite{Petrone2012,Shen2011}. This shows that the fractal nature of graphlocons does not significantly alter the electronic properties of graphene, despite the importance of the edge in quantum Hall physics.

\section{Conclusion}

In summary, we have synthesized large, highly dendritic graphene islands named graphlocons by CVD inside a copper enclosure. %Optical inspection emphasized the seeding role of copper impurities and morphological defects and Raman spectroscopy confirmed the presence of small bilayer domains in the middle of large monolayer graphene.%
By comparing the island shape evolution of graphlocons to other types of large island growths, we showed and quantified the distinct morphology of graphlocons. Based on this analysis we explained the formation of dendrites in CVD-grown graphene as the result of the competition between carbon attachment and diffusion along the graphene island in a surface-diffusion limited growth regime. Graphlocons were transferred onto SiO$_2$/Si, electrically contacted and a hole mobility as high as 6300 cm$^2$V$^{-1}$s$^{-1}$ was extracted. Similar Hall mobility values were found and magnetotransport measurements displayed well-developed QHE as well as strong SdHO. These observations all demonstrate the high quality of graphlocons and their potential for graphene-based electronics.

\begin{acknowledgement}

The author thanks Richard Chromik for the Raman spectrometer, Richard Talbot and Robert Gagnon, and the staff of the McGill Nanotools Microfab for the technical support. This research was supported by NSERC and FQRNT.
shown in this document.

\end{acknowledgement}

\begin{suppinfo}

\subsubsection{Other Growth Methods}

In addition to the growth method described in the article, we used two other growth techniques that are known for growing large graphene crystals by CVD on copper. The results of these growths are shown in \ref{fgr:growthmech}a. Following the method described by Ref.~\citenum{Zhang2012a}, the vapor trapping growth was performed by placing a piece of copper foil inside a small quartz tube (\ref{fgr:othergrowths}a and b). The foil was first annealed at 1050$^{\circ}$C for 30 min in 50 mTorr of H$_2$ flowing at 7 sccm. Graphene islands were synthesized at 1050$^{\circ}$C at a pressure of 200 mTorr, using a 1 sccm CH$_4$ flow and a 12.5 sccm H$_2$  flow. 

The square-island growth (\ref{fgr:othergrowths}c and d) was achieved by employing large H$_2$/CH$_4$ ratio as reported by Ref.~\citenum{Wang2012}. The copper foil was first annealed during 30 minutes in a flow of 10 sccm H$_2$ at 1050$^{\circ}$C. The growth was then performed at 1050$^{\circ}$C in a flow of  50 sccm H$_2$ and 1 sccm CH$_4$ at a pressure of 770 mTorr. After the growth the sample was cooled rapidly to room temperature under H$_2$ flow.

\begin{figure}
 \centering
\includegraphics[width=\columnwidth]{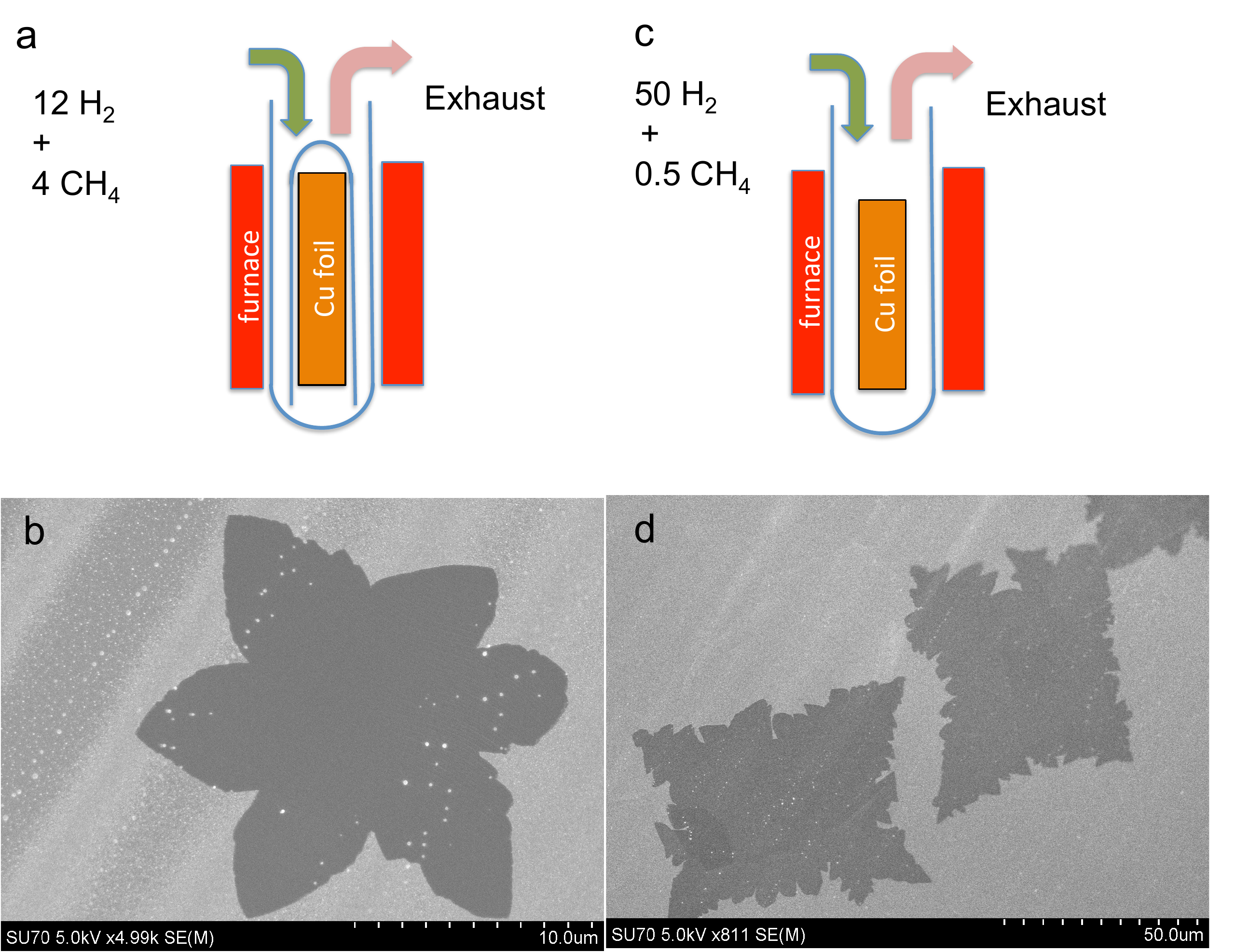}
  \caption{Other growth methods employed to grow large graphene islands. (a) Schematic of the vapor trapping setup with a copper foil inserted in a small inverted quartz tube and (b) SEM picture of a typical graphene island synthesized with this method. (c) Schematic of the square-island growth and (d) SEM picture of the synthesized islands.    }
  \label{fgr:othergrowths}
\end{figure}

\subsubsection{High Temperature Growth}
 
Growths were performed at a temperature of $\sim$1080$^{\circ}$C. Copper foils were first pre-treated using acetone, acetic acid and isopropanol. The copper foil was then annealed for 65 minutes in a flow of 75 sccm H$_2$ at a pressure of 650 Torr. Following the annealing process the pressure was lowered to $\sim$130 Torr and a flow of 0.15 sccm CH$_4$ was introduced for 30 minutes, while the H$_2$ flow was maintained at 75 sccm.
After the growth the sample was removed from the oven and allowed to cool rapidly to room temperature under H$_2$ flow.
\begin{figure}
\centering
\includegraphics[width=10cm]{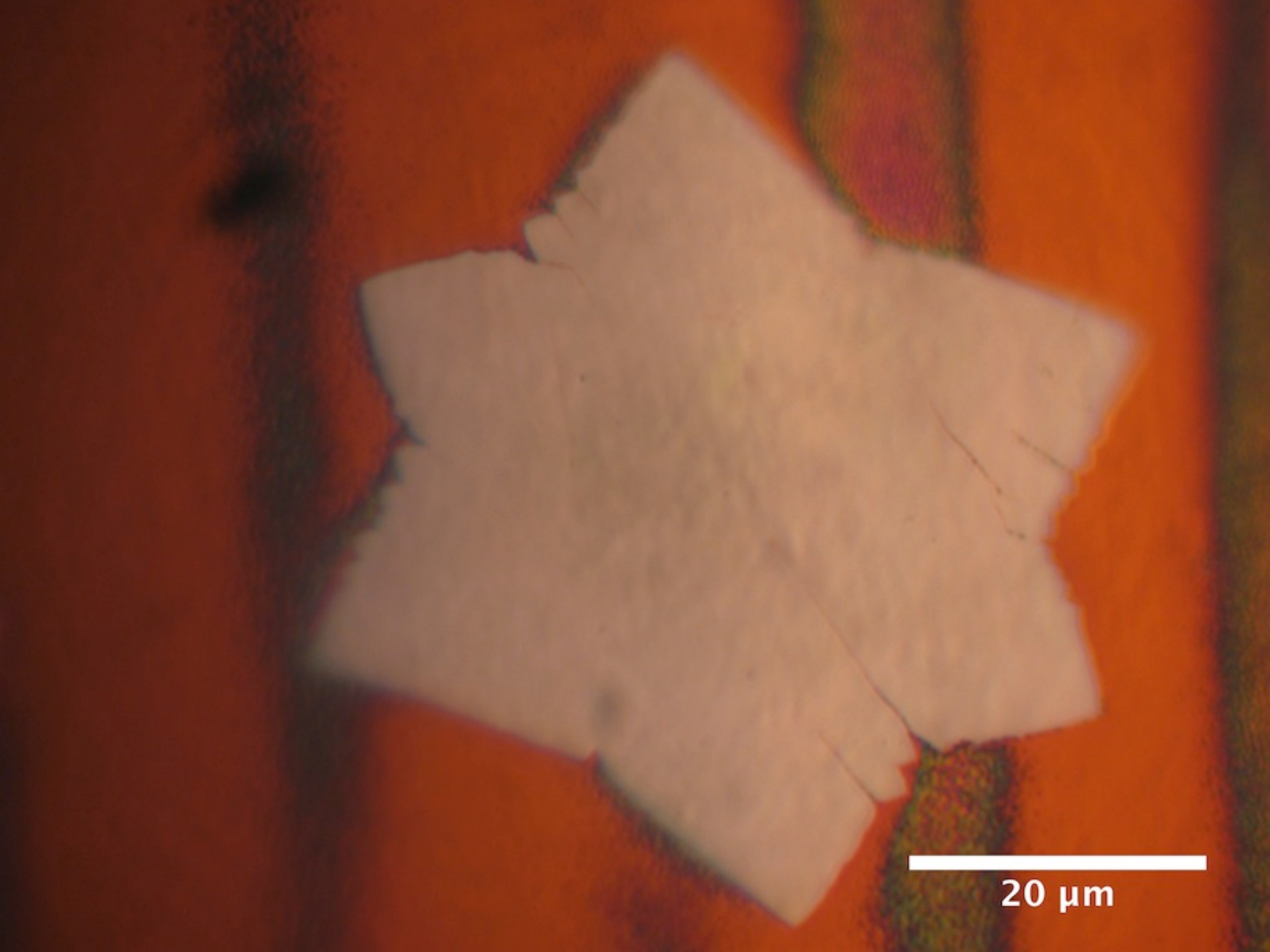}
\caption{Optical micrograph of graphene crystal grown using a high pressure, high temperature CVD process. The dark colored region corresponds to oxidized Cu}
\label{fgr:tmm3}
\end{figure}

\end{suppinfo}

\bibliography{graphlocon}

%% graphical table of contents.
%%%%%%%%%%%%%%%%%%%%%%%%%%%%%%%%%%%%%%%%%%%%%%%%%%%%%%%%%%%%%%%%%%%%%

\end{document}